\documentclass[aps,prl,preprint]{revtex4-1} 
\pdfoutput=1
\usepackage{graphicx}     
\usepackage{amssymb, amsmath}      
\DeclareMathSizes{7}{11}{3}{3}  
\usepackage{textpos}  
\usepackage{hyperref}
\hypersetup{colorlinks=true, citecolor=blue, urlcolor=blue, linkcolor=blue}   
\begin{document}
\begin{textblock*}{\textwidth}(0cm,0cm)
{{\it This is the accepted author version of the article published in} Physical Review B vol. 88, 195144 (2013) [$\copyright$ 2013 American Physical Society]. The version of record is available from the publisher's website from the link: \url{https://doi.org/10.1103/PhysRevB.88.195144}}
\end{textblock*}  
\vspace{4cm}       
  
\title{Extended slow-light field enhancement in positive/negative-index heterostructures} 
\author{S. Foteinopoulou$^{1, *}$ and J. P. Vigneron$^{2, \dagger}$}
\affiliation{\it $^{1}$ School of Physics, CEMPS, University of Exeter, Exeter, United Kingdom\\
\it $^{2}$ Facultes Universiterais Notre dam de la Paix (FUNDP), Namur, Belgium}
\begin{abstract}   
We present a bi-waveguide paradigm composed of joined Positive-Index-Material (PIM)/Negative-Index-Material (NIM) slabs, demonstrating ultra-slow light propagation stemming from the competing propagation disposition in the PIM and NIM regions. We report for the first time a mesoscopic extended electromagnetic (EM) enhancement covering regions of the order of the free space wavelength, enabled by the slow-light mode in our system. Our dynamic numerical results are consistent with our developed theoretical model, predicting an EM energy accumulation reminscent of a charging capacitor. Our analysis reveals that spatial compression is not a requirement to EM enhancement in slow-light systems and stresses on the merits of high coupling efficiency, strong temporal compression, monomodality and modal index bandwidth, -all present in our proposed paradigm. Furthermore, we show that the heterostructure waveguide mode is an extra-ordinary entity with a unique energy velocity, that is opposite to the Poynting vector in one of the participant waveguides. We believe these results will inspire new slow-light platforms relevant to the collective harvesting of strong light-matter interactions.    
\\
{PACS numbers:  81.05.Xj, 78.67.Pt, 42.25.-p, 42.82.Et} 
\end{abstract}      
\maketitle       
\par 
\section{I. Introduction} 
\par
Slow light is attracting an increasing attention in the last few years owing to its tremendous applications potential, as for example in all-optical chips \cite{krauss1, krauss2, povinell} and optical memories \cite{krauss2}. Slow light has furthered the frontiers of current optics, with many works reporting on strong light-matter interactions that manifest themselves as an enhancement of optical gain \cite{krauss1,baba}, Raman signal \cite{fan1} or non-linear optical processes \cite{krauss1,baba}, such as third-harmonic generation\cite{eggleton}. Strong light-matter interactions are associated with high-intensity electromagnetic (EM) fields. Intensity enhancing platforms have been pursuit vigorously in the recent years, which involve a spatial compression of the impinging wave into deep-subwavelength volumes. Typically, spatial light compression is facilitated by a resonant \cite{vidal, schatz, sfopex, schuller}  interaction between light and  a structured material -although a non-resonant\cite{nonres} scheme has also been recently demonstrated. 
\par
Slow-light physics is opening an alternate avenue where field enhancement may occur as a result of temporal compression \cite{krauss1, krauss2}, thus allowing it in principle to extend over mesoscale areas of the order of the free space wavelength. This can facilitate a collective harvesting of light-matter interaction, which is important when a strong signal yield from non-linear optical processes is desired. With the exception of resonant atomic systems \cite{khurgin}, all other slow-light paradigms  \cite{krauss2} seem promising in terms of field enhancement, which has been indirectly evidenced by observing a dependent process \cite{fan1, eggleton, marin}. However, a deeper understanding is still lacking with respect to which are the key controlling attributes pertinent to the dynamics of EM-energy enhancement in slow-light platforms. Also, there seems to be a lack of a clear consensus over the role of spatial compression in slow-light systems \cite{krauss2, baba}.     
\par  
Different schemes have been employed to control the light's dispersion and thus speed of information transfer in man-made architectures \cite{eit1, eit2, eit3, krauss1, krauss2, krauss3, krauss4, baba, baba2, marin, sridhar, karalis}. An interesting scheme was proposed in 2006 by C. Vandenbem et al. \cite{osapro} that joins together a positive refractive index medium (PIM) with a negative refractive index medium (NIM)\cite{Veselago} into a heterostructure waveguide \cite{engheta1}, shown to exhibit a flat photonic dispersion\cite{osapro}. Other NIM-based waveguides had been also reported by Shadrivov et al. \cite{andrei} and Tsakmakidis et al. \cite{kosmas}, but with the energy being guided through the negative index material only, evanescently decaying outside. In the system of Ref. \cite{kosmas} a slow light behavior leading to rainbow trapping via an adiabatic taper was predicted with frequency domain calculations. Given the on-going rapid development of photonic metamaterials \cite{soukr,shalr}, these NIM-based waveguides seem highly promising as slow-light platforms. However, the salient dispersion features relevant to an achievable ultra-high slow down factor, as well as the related unconventional propagation characteristics within a wave-optics picture have not been identified. Furthermore, the peculiar dynamics of energy accumulation in these systems, and its relation to spatial and temporal compression is certainly still outstanding. It should also be noted that efficient coupling to slow-light modes is not a trivial matter, and is the subject matter of current on-going research \cite{slcoup1, slcoup2}. 
\par 
In this paper, we propose a particular positive-index-medium (PIM)-negative-index-medium (NIM)  heterostructure as a composite bi-waveguide paradigm and discuss its extra-ordinary dispersion features in Sec. II. In Sec. III we present time-domain simulations that verify the existence of the trapped slow mode, -predicted by the frequency domain simulations of Sec. II- and also demonstrate an associated high EM enhancement covering mesoscale areas, of the order of the wavelength. We further analyze the propagation characteristics of the slow mode and report for the first time an exotic guided wave having a Poynting vector that is opposite to the direction of energy propagation in one of the participant waveguides in Sec. IV. These dynamic observations further stress on the importance of a large modal index bandwidth that we discuss in Sec. V. Furthermore, in Sec. VI we study the EM energy accumulation for the slow-mode in time-domain. We present an analytical model that underpins the observed dynamic EM energy accumulation, which reveals the participating roles of spatial and temporal compression. This model identifies the prominent features of our  prototype's band dispersion that are responsible both for observing ultra-slow light and achieving an extended high field-enhancement. Finaly we present our conclusions in Sec. VII. 
\par
\section{II. The NIM-PIM bi-waveguide paradigm}
\par
In Fig. 1(a) we show schematics of the bi-waveguide paradigm, with its geometric characteristics designated (we assume an infinite extend in the $x-$ and $z-$ directions). We take a dielectric medium for the PIM ($\varepsilon_2=4.0$, \hspace{0.5mm} $\mu_2=1.0$) and consider a homogeneous left-handed medium (LHM) \cite{Veselago} for the NIM, similar to the one in Refs. \cite{kirby, Pendry, rabia}. In particular we take:
\begin{equation}
\varepsilon_1= \mu_1=1-\frac{\omega_p^2}{\omega^2},
\end{equation}
with $\omega_p=2 \pi \cdot 308\cdot 10^{12} \hspace{0.1mm} rad/s$ and $\omega$ representing the frequency of the EM wave. We investigate the $H-$polarized composite guided modes, i.e. light is guided within both the NIM and PIM layers of the bi-waveguide and has evanescent tails outside, with magnetic field along the $z-$ direction. Our design prototype has $\textrm d_1=1702.5$ nm and $\textrm d_2=2837.5$ nm. 
\par
By assuming guided waves in both NIM, and PIM regions and evanescent tails outside [see also Appendix I] we obtain the dispersion relation for the composite NIM-PIM bi-waveguide guided mode -- $\omega(k_{||})$, with $k_{||}$ being the wavevector along the $x$-axis. This is given by:
\par
\begin{equation}
\frac{(z_1+z_2) \textrm{cos}\hspace{0.01cm} OP + (1-z_1z_2) \textrm{sin}\hspace{0.01cm} OP}{(z_1-z_2) \textrm{cos}\hspace{0.01cm}\Delta OP + (1+z_1z_2) \textrm{sin} \hspace{0.01cm} \Delta OP}=\frac{z_1-z_2}{z_1+z_2},
\end{equation}
\par
where $OP=k_{1y} \textrm d_1 +k_{2y} \textrm d_2, \hspace{2mm}$ $\Delta OP=k_{1y} \textrm d_1-k_{2y} \textrm d_2$ and $z_1=\frac{k_{1y}}{\varepsilon_1 k_y}$, \hspace{3mm} $z_2=\frac{k_{2y}}{\varepsilon_2 k_y}$. $k_y$ represents the decaying wavevector outside the bi-waveguide while $k_{1y}$ and $k_{2y}$ represent the wavevectors along the $y$-axis inside the negative-index part and positive-index part of the bi-waveguide respectively. i.e.:
\par
\begin{equation}
k_y=\sqrt{k_{||}^2-\frac{\omega^2}{c^2}},
\end{equation}
\begin{equation}
k_{1y}=\sqrt{\varepsilon_1 \mu_1 \frac{\omega^2}{c^2}-k_{||}^2}, 
\end{equation}
and
\begin{equation}
k_{2y}=\sqrt{\varepsilon_2 \mu_2 \frac{\omega^2}{c^2}-k_{||}^2},   
\end{equation} 
\\
with $c$ being the speed of light.
\par
We solve Eq. (2) as a transcedental equation and show the result for the photonic dispersion in Fig. 1(b) with a solid line. Notice, the bi-waveguide mode transitions from a region of a positive band-slope to a region of a negative band-slope  with increasing $k_{||}$. Essentially, this implies that the guided mode transitions from being forwards to being backwards \cite{tretyakov} by passing through a regime with near-zero $d \omega/dk_{||}=\textrm v_g$  implying near-frozen light [see also dashed line in Fig. 1(c)]. Now, this by itself may not be striking and occurs also in systems where the EM energy is guided through the LHM core only\cite{andrei}. However, what is striking is that the near-zero $\textrm v_g$ regime extends over a much larger $k_{||}$ interval, that we refer to as modal-index bandwidth from now on \cite{modalindex}. This extra-ordinary feature was observed before in photonic-crystal waveguides \cite{krauss2, krauss3}, but in these system does not routinely come with monomodality. Cross-modal talk directs EM energy away from the slow-light channel thus being a major hurdle in the performance of photonic-crystal based platforms. On the other hand, monomodality is an inherent advantage in these metamaterial bi-waveguides \cite{engheta1} and flat-bands with a large modal-index bandwidth can be easily tailored with a proper choice of  $\textrm d_1$ and $\textrm d_2$ \cite{osapro}. We note that the key importance of these essential characteristics has not been discussed in other NIM-based slow-light platforms \cite{kosmas}, where the light slow-down factor has not been attested in the time-domain \cite{kirby}.   
\par  
We highlight in Fig. 1 the modal-index bandwidth for near-frozen light with the dark shaded region, which transitions to the left and right to a $k_{||}$-region of forwards and backwards faster light respectively. Notice, that the composite guide mode is on the right side of the air-lightline (dashed-line)  but on the left side of the lightlines in the NIM (dotted line) and in the PIM (not seen since much further to the right of figure bounds). This signifies the guided-wave nature of the mode within both the PIM and NIM parts of the bi-waveguide heterostructure. We also stress that indeed the group velocity of our system [dashed line in Fig. 1(c)] portrays the velocity of energy propagation, $\textrm v_e$ \cite{Veselago} (See Appendix I). [solid line in the same figure]. 
\par
\section{III. Dynamic behavior of light in the NIM-PIM bi-waveguide}
\par
In the following, we will study the dynamic behavior of light in our proposed paradigm. For this purpose we employ the Finite-Difference-Time-Domain (FDTD) method \cite{taflove, pml} -- a proven excellent method to capture the time-evolution of EM waves while propagating in space even within negative index materials \cite{rabia, ziolkowski}. To excite the composite guided mode in the bi-waveguide heterostructure we employ in FDTD an Attenuated Total Reflection (ATR) set-up \cite{voulaback, voulaatr2} in the Otto configuration \cite{otto} as depicted in Fig. 2(a) \cite{voulaback}. The refractive index of the prism, $\textrm n_{\textrm{atr}}$, varies as required to yield the desired $k_{||}$ value \cite{voulaatr1}. In this manner, we can select from the entire band of Fig. 1(b) a specific band-region centered around modal index $c k_{||}/\omega$ \cite{modalindex}, with bandwidth that is inversely proportional to the beam waist of the incident Gaussian beam, depicted in the schematics of Fig. 2(a) with its magnetic field polarized out-of the prism plane. 
\par     
We have confirmed numerically the band dispersion, by considering a pulsed impinging signal and recording the Fourier-transformed spectral response of properly placed detectors above the prism and laterally within the bi-waveguide. We  show our results in Fig. 1 as filled circles. We highlight there three particular modes that we designate with open diamonds and name as M1, M2, and M3. These have respectively a positive, near-zero and negative band slope implying a forward, slow-light and backward propagation in each case. 
\par
With a quasi-monochromatic excitation \cite{voulaprlneg} at the relevant frequency in FDTD we confirm the forwards and backwards lateral propagation as we can see in Figs. 2(b) and (d) respectively, where the time-averaged electric-field intensity is shown cropped around the waveguide area. [See also Figs. 3(a) and 3(c) respectively, where the ATR prism excitation area is included]. Now, mode M2, seen in Fig. 2(c) [or Fig. 3(b)],  is mostly forwards propagating, but we see some EM energy escaping to the opposite side, due to the unavoidable non-zero $k_{||}$ bandwidth of the impigning Gaussian source. What is striking is the much higher intensity enhancement with respect to the intensity of the input beam that is associated with this mode.  
\par
\section{IV. The interlocking of the NIM and PIM sub-modes with counter-directional Poynting vectors}
\par 
We proceed in analyzing further the curious characteristics of the composite slow-light mode in the heterostructure bi-waveguide. For this purpose, we launch in the FDTD simulation a narrow-bandwidth pulsed signal \cite{mono} and observe the Poynting vector within each of the joint waveguides. We monitor the $x$-component of the Poynting vector at different lateral locations, at the right-side of the prism for the forwards modes, i.e. modes M1 and M2, and at the left-side of the prism for the backwards mode, i.e. mode M3. Let, $\textrm D_{\textrm{det}}$ represent the distance between the lateral line detectors and the relevant side of the prism edge. For each lateral location, $\textrm D_{\textrm det}$, one line detector is placed within the NIM part of the bi-waveguide and one line detector is placed within the PIM of the bi-waveguide. Then, for each time step, the $x$-component of the Poynting vector $\textrm S_\textrm x$, is integrated along the respective line detectors in the NIM and PIM parts of the bi-waveguide. We represent this integrated value as $\textrm S_{\textrm x,\textrm{int}}$. In Fig. 4 we show the result for mode M2 and $\textrm D_{\textrm det}=2.27$ microns. We show the respective results for modes M1 and M3 in Fig. 5.
\par
We find that the pulse propagates laterally, i.e. along the x-direction, consistent with a wave-optics picture. A geometric optics picture, previously employed to explain slow-light propagation in Ref. \onlinecite{kosmas}, clearly fails, as it implies a significant Poynting vector component along the $y-$ direction, which we did not observe here in our simulations. We observe in FDTD a truly astonishing phenomenon, where Poynting vector is antiparallel to the direction of EM energy propagation in one of the participant waveguides. Specifically, the Poynting vector consistently points towards the $-x$ direction in the NIM layer and towards the $+x$ direction in the PIM layer for all cases. Indeed, in all cases the sub-modes within each of the participant waveguides interlock together. They move jointly, as an entity, towards the same direction that is determined by the band slope of the dispersion relation, $\omega(\textrm k_{||})$, despite the counter-directional relation of the Poynting vector within each layer of the heterostructure waveguide.   
\par
The behavior of the guided waves within each sub-guide as an entity can be further evidenced by observing in the FDTD simulation the arrival time \cite{peatross} of the pulsed signal at different locations at the right side of the prism (for modes M1 and M2 that are forwards) or the left side of the prism (for mode M3 that is backward). The pulse arrival time within the NIM and PIM parts of the bi-waveguide, at the different lateral locations, will be calculated from the detected $\textrm S_{\textrm x,\textrm{int}}$ with time. Normally, in experiments or simulations the arrival time of a pulse is determined by the peak of the pulse. However, this is a rough method and can entail significant errors for systems where significant pulse broadening occurs during propagation. This is particularly true for slow light systems. J. Petross et al. \cite{peatross}, have proposed an accurate measure for the pulse arrival time for non-standard systems that we employ here. Thus, we calculate the pulse arrival time, $t_{\textrm{arr}}$, at a certain lateral line detector, $\textrm D_{\textrm{det}}$ from the following relation:
\par
\begin{equation} 
t_{\textrm{arr}}=\frac{\int \limits_{0}^{t_{\textrm{sim}}} t \textrm S_{\textrm x,\textrm{int}}(\textrm D_{\textrm{det}}, t) dt}{\int \limits_{0}^{t_{\textrm{sim}}} \textrm S_{\textrm x,\textrm{int}}(\textrm D_{\textrm{det}}, t) dt},
\end{equation}
with $t$ being the time-instant where $S_{\textrm x,\textrm{int}}$ is monitored at the line detectors in the NIM and PIM part of the bi-waveguide at the $\textrm D_{\textrm{det}}$ lateral location. Also, $t_\textrm{sim}$ is the total simulation time. Note, since we have different line detectors placed in the participant waveguides of the heterostucture, arrival time is recorded separately in the NIM and PIM parts of the bi-waveguide. 
\par 
We plot the results for the arrival time at the different lateral detector positions $\textrm D_{\textrm{det}}$, in Fig. 3 in panels (a), (b) and (c) for the modes M1, M2 and M3 respectively. The filled circles represent the arrival times calculated in the NIM part of the bi-waveguide. Conversely, the open circles represent the arrival times calculated in the PIM part of the bi-waveguide. The excellent agreement between the two ascertains further our previous observation that the composite mode is an entity, and EM energy propagates with the same speed within the entire waveguide extend, both within the NIM and PIM parts.
\par
Now, we can take a linear fit in each case and from this we obtain the energy velocity of the composite mode. We express this in terms of the speed of light $c$ in each case and show it in the respective figures. Indeed, the observations for the energy velocity within the NIM and PIM part of the bi-waveguide are in excellent agreement, further attesting that the EM energy of the composite mode propagates with a unique speed in all parts. We have expressed the results in fractional form, so the slow down factor is immediately evident. We find a slowdown factor of $\sim 16$, for mode M1,  $\sim 303$ for mode M2 and $\sim 9$ for mode M3. 
\par
To compare with the theoretical predictions from the dispersion relation we enlist from Fig. 1 the slow-down factor, that correspond to the central $\textrm k_{||}$ of the impinging Gaussian beam; these are  $\sim 13$, for mode M1,  $\sim 105$ for mode M2 and $\sim 9$ for mode M3. We observe an excellent agreement for modes M1 and M3 but we find that mode M2 is slower than expected. This is because the finite spatial extend of the impinging beam, implies a $\textrm k_{||}$ span within a band $\Delta \textrm k_{||}$. Thus in practical situation one observes the average speed from the mode contributions within this  $\Delta \textrm k_{||}$ band. For mode M2, this ended up yielding even a higher slow-down factor than the one predicted for the central $\textrm k_{||}$ value. 
\par
\section{V. Slow light propagation and modal index bandwidth}
\par
Our analysis in Sec. IV suggests that the overall speed of the EM energy propagation is affected by the finite spatial extent of the impinging Gaussian beam implying contributions from within a $\Delta \textrm k_{||}$ band. This further highlights the particular advantage of having a wide modal index span over which the composite bi-waveguide mode is near-frozen, as we have seen in Fig. 1. For example, let's consider a comparative design with $\textrm d_1$=2270 nm and 
$\textrm d_2$=3405 nm. We show the dispersion relation $\omega(k_{||})$ and corresponding energy velocity (in units of speed of light $c$) for the latter design in Figs. 7(a) and 7(b) respectively. In  order to be able to easily compare the result with the system of Fig. 1 we also show the respective dispersion and energy velocity as dashed lines. Evidently the band for the comparative design is not as flat. A near-frozen mode exists only for a very narrow $k_{||}$ band.
\par 
We further investigate in FDTD, the implications of the lack of existence of a wide modal-index bandwidth. We set the ATR prism properties to excite in the comparative design the near-frozen mode of $ck_{||}/\omega=1.12$, as predicted from the photonic dispersion of Fig. 7. Yet what we observe (see Fig. 8) is an amphoteric propagation. This emanates from the strong contributions from the adjascent backwards and forwards modes, as the near-frozen mode has a narrow modal index bandwidth in the comparative waveguide. The latter cannot be avoided due to the finite spatial extend of any realistic excitation beam. These results suggest that
near-frozen light modal-index bandwidth is the key feature of merit, when considering the design characteristics of the bi-waveguide.
\par
\section{VI. EM energy accumulation in the bi-waveguide} 
\par
Now that we have verified a large slow-down factor for the M2 mode, we try to understand its relation to the observed intensity enhancement seen in Fig. 2(c) [or 3(b)], which spans over the entire waveguide width, $d\sim3\lambda_0$, with $\lambda_0$ being the free space wavelength. Since the waveguide extent in the $z$-direction is infinite, implying that in practice the waveguide can be arbitrary wide in this direction, our results essentially suggest an extended intensity enhancement covering a large mesoscale area, larger than $\lambda_0^2$. This to our knowledge constitutes a first report of this capability, as typically intensity enhancement is restricted within deep-subwavelength regions \cite{vidal, schatz, sfopex, schuller, nonres}.   
\par
To understand this further, we excite the M2 slow mode with a quasi-monochromatic wave and monitor the time-evolution of the EM energy density, spatially averaged over a central region of the bi-waveguide lying directly below the ATR prism. We take the time-average of this quantity within each wave-period T, $<\textrm U>$, and normalize it with the time-averaged EM energy density of the source $\textrm U_0$. The result at each period T, shown in Fig. 9(b), implies an energy accumulation with an almost exponential time-response reaching a two-hundred-fold EM energy density enhancement at steady-state. 
\par
Let's consider a simple crude model, depicted in Fig. 9(a), to explain the energy accumulation dynamics. EM energy, $\mathcal E$, is fed into the bi-waveguide from above via the ATR prism, at a faster rate than its relaxation sideways in the $+x$ direction [M2 is a forwards mode] \cite{counter}, because of the ultra-low energy velocity of the M2 mode. In the following with brackets $< >$ we denote the spatial averaged quantities of the energy density, in an area $\textrm L_{\textrm w} \textrm d$ in the central part of the bi-waveguide below the ATR prism, where d represents the total width of the bi-waveguide [see schematics of Fig. 9(a)]. With $\textrm L_{\textrm w}$ we will denote the beam waist of the evanescent wave illumination emanating from the ATR prism, which is close to the Gaussian impinging wave's beam waist. Moreover, time-averaged quantities, within a wave period, T are assumed for the EM energy densities, U.  Note, we have translational symmetry in the $z$ direction, where the guide is assumed infinite. We focus on a part of the guide with a width $L_{\textrm{per}}$ along $z$. We will see that this chosen part can be arbitrary and does not influence the result. Then, within $\Delta T$, the EM energy that gets fed, into the waveguide would be:\\
\begin{equation} 
\Delta \mathcal E_{\textrm{feed}}=F_c I_0 \Delta T \textrm L_{\textrm w} \textrm L_{\textrm{per}},
\end{equation}
where $F_c$ represents the coupling efficiency and $I_0$, the intensity of the impinging Gaussian Beam, thus: 
\begin{equation}
\Delta \mathcal E_{\textrm{feed}}=F_c \hspace{0.3mm} c \hspace{0.3mm} {U_0} \Delta T \textrm L_{\textrm w} \textrm L_{\textrm{per}},
\end{equation}
where $U_0$ is the energy density of the impinging beam. Then the energy that get's relaxed sideways within $\Delta T$ would be:
\begin{equation}
\Delta \mathcal E_{\textrm{rel}}=<S_x> d \hspace{1mm} \textrm L_{\textrm{per}} \Delta T= \textrm v_e <U> d \hspace{1mm} \textrm L_{\textrm{per}} \Delta T
\end{equation}
So, the increase in EM energy in $\Delta T$ would be, $\Delta \mathcal E_{\textrm{feed}}-\Delta \mathcal E_{\textrm{rel}}$. \\
This equals to $<\Delta U> d \hspace{1mm} \textrm L_{\textrm w} \textrm L_{\textrm{per}}$ yielding together with Eq. (8) and (9) that:
\begin{equation}
\frac{\Delta <U>}{\Delta T}=-\frac{\textrm v_\textrm e}{\textrm L_{\textrm w}}(<U>-U_M)
\end{equation}
where 
\begin{equation}
U_m=F_c \hspace{0.1cm} U_0 \frac{c}{\textrm v_\textrm e} \hspace{0.1cm} \frac{\textrm L_{\textrm w} }{\textrm d}.
\end{equation}
Eq. (10), with Eq. (11) yields:
\begin{equation}
<U>=U_0 \frac{c}{\textrm v_\textrm e} \hspace{0.1cm} \frac{\textrm L_{\textrm w} }{\textrm d} (1-e^{-\frac{\textrm v_\textrm e}{\textrm L_{\textrm w}}} t),
\end{equation}
where time, t, should be an integer multiple of period T. 
\par
Thus, the estimation of the feeding and relaxation rates leads to a time-response for the EM energy accumulation that is reminiscent of a charging capacitor, with the characteristic time $\tau$ being equal to ${\textrm L_\textrm w}/\textrm v_{\textrm e}$. Eq. (12) implies that spatial compression contributes a factor of ${\textrm L_\textrm w}/{d}$ to the overall EM energy enhancement, which is actually quite modest (about four) for the bi-waveguide paradigm. Temporal compression contributes a large factor for slow light modes equal to ${c}/{\textrm v_\textrm e}$. The results of Fig. 9(b) also imply a high coupling efficiency $F_c$, more than 60$\%$ with our proposed simple ATR-based coupling scheme.  We observe that the overall EM energy enhancement factor is not evenly spread within each sub-waveguide and is also larger than the corresponding electric field intensity enhancement (enhancement of $EE^{*}$). Specifically, we find the latter to be about a hundred in the NIM layer and about thirty in the PIM layer. Note also, that the characteristic time of the exponential EM energy growth, indirectly provides the light slow down factor  which we found to be with an exponential fit about 90 for the case of Fig. 9(b)\cite{slowdiff}.   
\par
We note that the comparative design yielded a much lower EM energy enhancement, as can be seen in Fig. 10. This is because additional energy relaxes from the faster, adjascent to the near-frozen mode, backward and forward modes, as any excitation beam inadvertendly includes contributions within a $\Delta k_{||}$ band. Eq. (12) simultaneously stresses the need for a wide modal-index bandwidth of the near-frozen mode but also for monomodality. Actually, it can be shown that the co-existence of any faster channel significantly hampers the maximum attainable EM energy enhancement.
\par
For example, suppose that light couples inside the composite waveguide with a total coupling efficiency $F_c$, but to two different modes that co-exist with respected weights $F_{c1}/F_c$ and $F_{c2}/F_c$, with $F_{c1}+F_{c2}=F_{c}$ and energy velocities ${v_{e1}}$, ${v_{e2}}$. Then the balance of energy fed and energy released sideways will lead to:
\begin{equation}
<U>=U_0 \frac{c}{\textrm v_\textrm e^{\textrm{eff}}} \hspace{0.1cm} \frac{\textrm L_{\textrm w} }{\textrm d} (1-e^{-\frac{\textrm v_\textrm e^{\textrm{eff}}}{\textrm L_{\textrm w}}} t),
\end{equation}
with
\begin{equation}
{\textrm v_\textrm e^{\textrm{eff}}}=\frac{{\textrm v_{\textrm e1}} F_{c1} + {\textrm v_{\textrm e2}} F_{c2}}{F_c}
\end{equation}
Eqs. (13) and (14) clearly imply that the maximum EM energy enhancement that can be attained is adversely affected by the co-existence of any faster channel. They also indicate that the respective saturation time to the maximum EM energy is smaller. This is consistent with what we observed in Fig. 10 for the energy accumulation for the comparative design of Fig. 7, where part of the impigning light couples to faster channels that are adjascent to the near-frozen mode.
\par
\section{VII Conclusions}
\par
In conclusion, we have presented a paradigm NIM-PIM heterostructure supporting an exotic composite guide mode, having an ultra-low energy velocity across a very broad modal-index bandwidth. We have observed in FDTD a most extra-ordinary propagation, where the sub-modes in each layer of the heterostructure interlock together and move jointly in the same direction despite the Poynting vector showing in the opposite direction in one of the layers. We verified numerically an efficient coupling to the slow mode, with speed about $c/300$, leading to exponential growth in the EM energy accumulation, reminiscent of a charging capacitor. Our findings suggest the possibility to achieve an extended electric field enhancement of the order of a 100, facilitated by the near-frozen waveguide mode of  ultra-wide modal index bandwidth. We therefore believe, this study will inspire new designs for slow-light platforms for the collective harvesting of strong light-matter interactions.
\par
\section{Appendix I: Energy velocity of the composite guided mode}
\par
The magnetic field distribution for the composite guided mode is given by:\\
$H_z$=$\begin{cases}
\textrm{A} e^{k_y (y + \textrm d_1)} \hspace{0.1mm} e^{i (k_{||} x- \omega t)} \hspace{5mm}  \textrm{for} \hspace{5mm} y< -\textrm d_1\\
(\textrm{B} \sin {k_{1y} y}+ \textrm{C} \hspace{0.5mm} \cos {k_{1y} y})\hspace{0.1mm} e^{i k_{||} x} e^{-i\omega t} \hspace{5mm}  \textrm{for} \hspace{5mm} -d_1 \leq y \leq \textrm 0\\
(\textrm{D} \sin{k_{2y} y}+ \textrm{E} \cos {k_{2y} y})\hspace{0.2mm} e^{i k_{||} x} e^{-i\omega t}\hspace{5mm}  \textrm{for} \hspace{5mm} 0 \leq y \leq \textrm d2\\
\textrm{F} e^{-k_y (y - \textrm d_2)}\hspace{0.5mm} e^{i k_{||} x} e^{-i\omega t} \hspace{5mm}  \textrm{for} \hspace{5mm} y> \textrm d_2\\
\end{cases}$\\
\\
where $y=0$ is taken at the PIM-NIM interface. The coefficients in the above $H_z$ field distribution are easily determined from the EM boundary conditions at the three interfaces (at $y=-\textrm d_1$, $y=0$ and $y=\textrm d_2$). Thus both the magnetic-field distribution and the corresponding electric-field distribution can be easily calculated for the composite guided mode. Then its energy velocity, $\textrm v_\textrm e$, along the $x$-axis [see schematics of Fig. 1(a)] can be given given by:
\begin{equation}
\textrm v_\textrm e=\frac{\bar S_x}{\bar U}=\frac{\frac{1}{8\pi} \int \limits_{-\infty}^{\infty} E_y H_z^{*} \, dy}{\frac{1}{16\pi}\int \limits_{-\infty}^{\infty} (\frac{\partial (\varepsilon \omega)}{\partial \omega} (E_x E_x^{*}+E_y E_y^{*})+\frac{\partial (\mu \omega)}{\partial \omega} H_z H_z^{*}) \, dy},\\
\end{equation}
where the expressions in the numerator and the denominator represent the time-averaged Poynting vector and the time-averaged energy-density in C.G.S. units, respectively, integrated along the finite dimension of the composite guide ($y$-axis) and outside to include the contributions from the evanescent tails in vacuum. Hence, the limits span from $-\infty$ to $\infty$  The general expressions for dispersive media \cite{Veselago} is taken for the energy density, with $\varepsilon$ being either $\varepsilon_1$ or $\varepsilon_2$ and $\mu$ being either $\mu_1$ or $\mu_2$, depending which region of the bi-waveguide $y$ lies within, or $\varepsilon=\mu=1$ for the vacuum region. Evidently, for the nondispersive material regions $\frac{\partial (\varepsilon \omega)}{\partial \omega}=\varepsilon$ and$\frac{\partial (\mu \omega)}{\partial \omega}=\mu$.   
\par
 
\newpage
{\bf List of Figures}
\par
{\bf Fig. 1}{(Color online) (a) Schematics of the PIM-NIM bi-waveguide (b) Photonic dispersion (theory) of the bi-waveguide's guide mode (solid black), with the dashed (dotted) line representing the lightlines in vacuum (NIM) (Not seen here is the lightline in the PIM at much larger $k_{||}$). The solid circles represent the corresponding numerical time-domain result, with the empty diamonds designating three characteristic cases labeled as M1, M2 and M3. Both the frequency $\omega$ and the $x$-component of the  wavevector, $k_{||}$, are scaled to be dimensionless ($\omega_p$ defines the material properties of the NIM).  The dark highlighted area tags the near-frozen light regime that merges to a regime with slow-backwards (slow-forwards) light to the right (left) (lighter highlighted area). The corresponding energy and group velocities, both scaled with the speed of light $c$, are also shown in (c) as solid and dashed lines respectively.}
\par
{\bf Fig. 2}{(Color online) (a) Schematics of the numerical FDTD ATR experiment set-up. In (b), (c) and (d) the FDTD result for the time-averaged electric field intensity, I, is shown for modes M1, M2 and M3 respectively. The domain is cropped to depict I only in the vicinity of the bi-waveguide, --extending within the horizontal solid lines of the figure. The vertical lines delimit the bi-waveguide region that lies directly below the ATR prism while the dotted lines designate the NIM-PIM boundary. Note, the colormap scale is logarithmic.}
\par
{\bf Fig. 3}{(Color online) Same as in Fig. 2(b)-(d) but with the ATR excitation region included. The triangle represents the ATR prism, used to excite the guided modes, while the solid lines represent the bi-waveguide limits.}
\par
{\bf Fig. 4}{(Color online) The  interlocking of the sub-modes within each layer of the heterostructure waveguide, observed in FDTD for mode M2 via the $x$-component of the Poynting vector monitored at the right of the ATR prism edge. The dark-colored (light-colored) line represents the result integrated over the $y$-extent of the NIM (PIM) part of the bi-waveguide. The Poynting vector is given in arbitrary units, while time is expressed in terms of the period, T of the central frequency $\omega_0$, of the pulsed signal. Clearly, the sub-modes move together towards the $+x$ direction, despite the Poynting vector showing towards the $-x$ direction in the NIM layer. [see schematics on the right of the figure indicating the wavevector ${\bf k_{||}}$ and energy velocity ${\bf v_e}$ of the composite guided mode, as well as the Poynting vector {\bf S} in each of the sub-waveguides]}
\par
{\bf Fig. 5}{(Color online) Same as in Fig. 4 but for the cases of mode M1 in (a), and mode M3 in (b)}
\par
{\bf Fig. 6}{(Color online) Pulse arrival times in the NIM (PIM) part of the bi-waveguide in the different lateral detectors represented by filled (open) circles. The arrival time is given in terms of the period, while the lateral detector locations $\textrm D_{\textrm{det}}$ in terms of the wavelength that corresponds to the central frequency $\omega_0$ of the impinging pulse. The line fit in each case yields the energy velocity of propagation in each sub-waveguide.}
\par
{\bf Fig. 7}{(Color online) Photonic band dispersion in (a) and energy velocity in (b) are shown with a solid line for a comparative bi-waveguide design with $\textrm d_1$=2270 nm and $\textrm d_2$=3405 nm. The green dashed and dotted lines represent the relevant lightlines as in Fig. 1. To aid the comparison, we show the respective result for the original design of Fig. 1 with dashed lines.}
\par
{\bf Fig. 8}{(Color online) FDTD time-averaged intensity result, suggesting an amphoteric propagation in the comparative waveguide design of Fig. 7 for the near-frozen mode of $ck_{||}/\omega=1.12$.}
\par
{\bf Fig. 9}{(Color online) Energy accumulation dynamics in the bi-waveguide paradigm. (a) Feeding of EM energy from above via the ATR prism and its relaxation sideways in the bi-waveguide is depicted. (b) The solid line represents the FDTD result for the time-averaged EM energy density of the spatial average in the bi-waveguide region below the center of the ATR prism $<\textrm U>$ for mode M2, normalized with the time averaged energy density of the source $\textrm U_0$. The dotted line represents a fit from an analytical dynamical model in accordance to the picture in (a).} 
\par
{\bf Fig. 10}{(Color online) Same as in Fig. 9(b) but for the comparative waveguide of Fig. 7. As a result of the narrow modal bandwidth for the near-frozen mode, a smaller EM energy enhancement and quicker saturation in comparison with the design of Fig. 1 is observed.}
\clearpage
\newpage
\par 
\begin{figure}[!ht]
\begin{center} 
\includegraphics[angle=0,width=8.5cm]{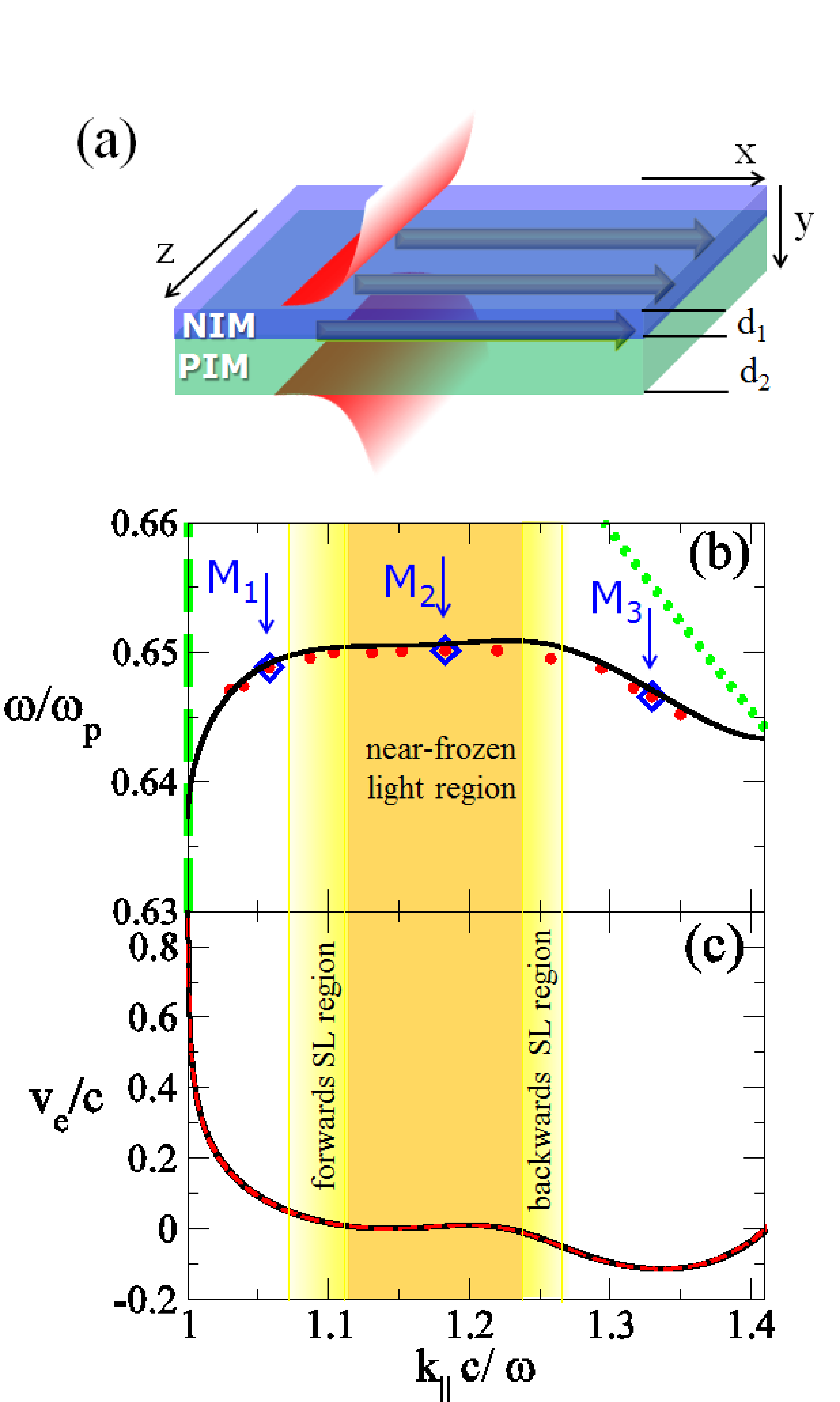}
\caption{(Color online) (a) Schematics of the PIM-NIM bi-waveguide (b) Photonic dispersion (theory) of the bi-waveguide's guide mode (solid black), with the dashed (dotted) line representing the lightlines in vacuum (NIM) (Not seen here is the lightline in the PIM at much larger $k_{||}$). The solid circles represent the corresponding numerical time-domain result, with the empty diamonds designating three characteristic cases labeled as M1, M2 and M3. Both the frequency $\omega$ and the $x$-component of the  wavevector, $k_{||}$, are scaled to be dimensionless ($\omega_p$ defines the material properties of the NIM).  The dark highlighted area tags the near-frozen light regime that merges to a regime with slow-backwards (slow-forwards) light to the right (left) (lighter highlighted area). The corresponding energy and group velocities, both scaled with the speed of light $c$, are also shown in (c) as solid and dashed lines respectively.}
\end{center} 
\end{figure}  
\clearpage
\newpage  
\par
\begin{figure}[!hbt] 
\begin{center} 
\includegraphics[angle=0,width=8.5cm]{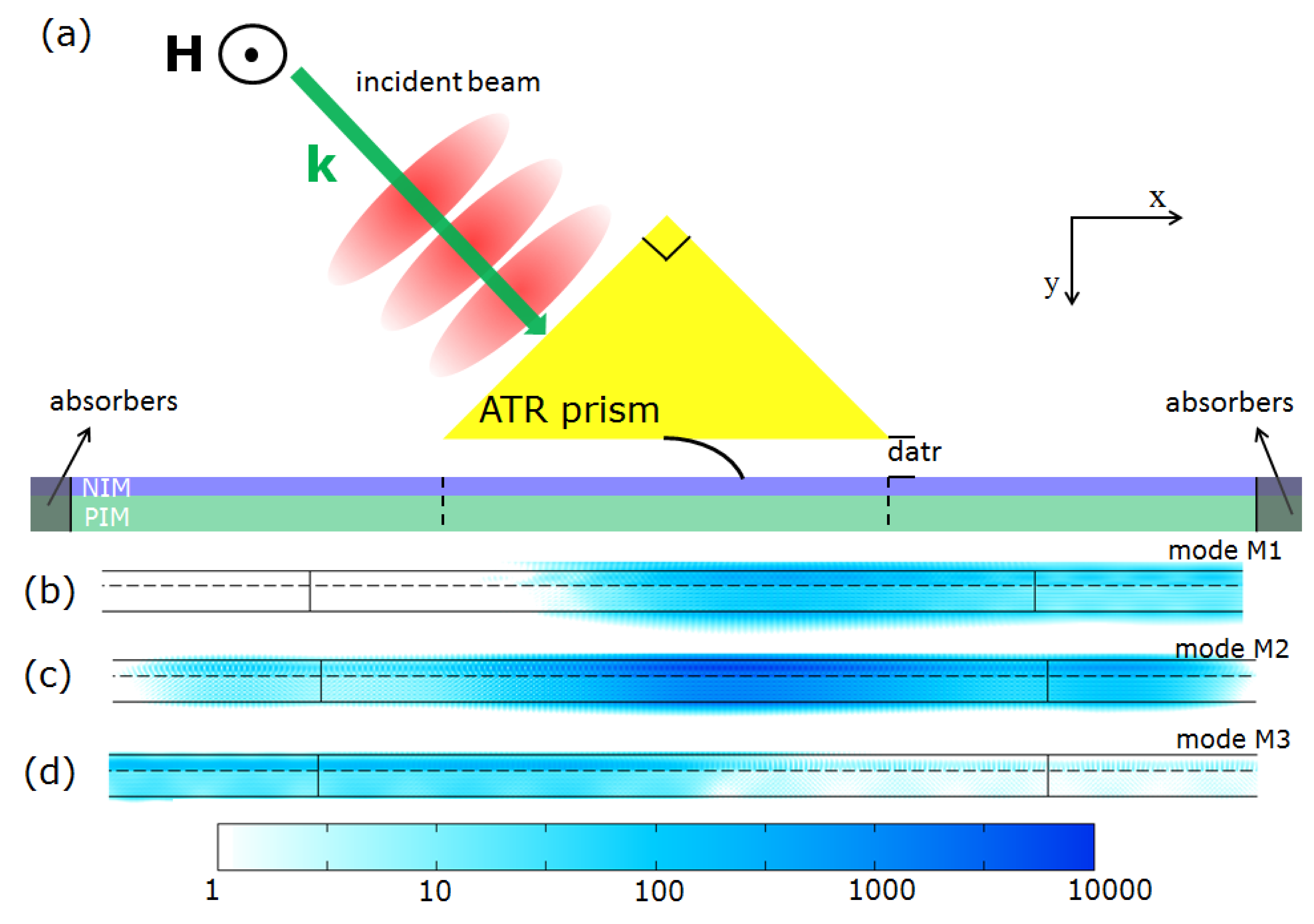}
\caption{(Color online) (a) Schematics of the numerical FDTD ATR experiment set-up. In (b), (c) and (d) the FDTD result for the time-averaged electric field intensity, I, is shown for modes M1, M2 and M3 respectively. The domain is cropped to depict I only in the vicinity of the bi-waveguide, --extending within the horizontal solid lines of the figure. The vertical lines delimit the bi-waveguide region that lies directly below the ATR prism while the dotted lines designate the NIM-PIM boundary. Note, the colormap scale is logarithmic.}
\end{center}   
\end{figure}  
\clearpage
\newpage
\begin{figure}[!hbt]
\begin{center} 
\includegraphics[angle=0,width=12.0cm]{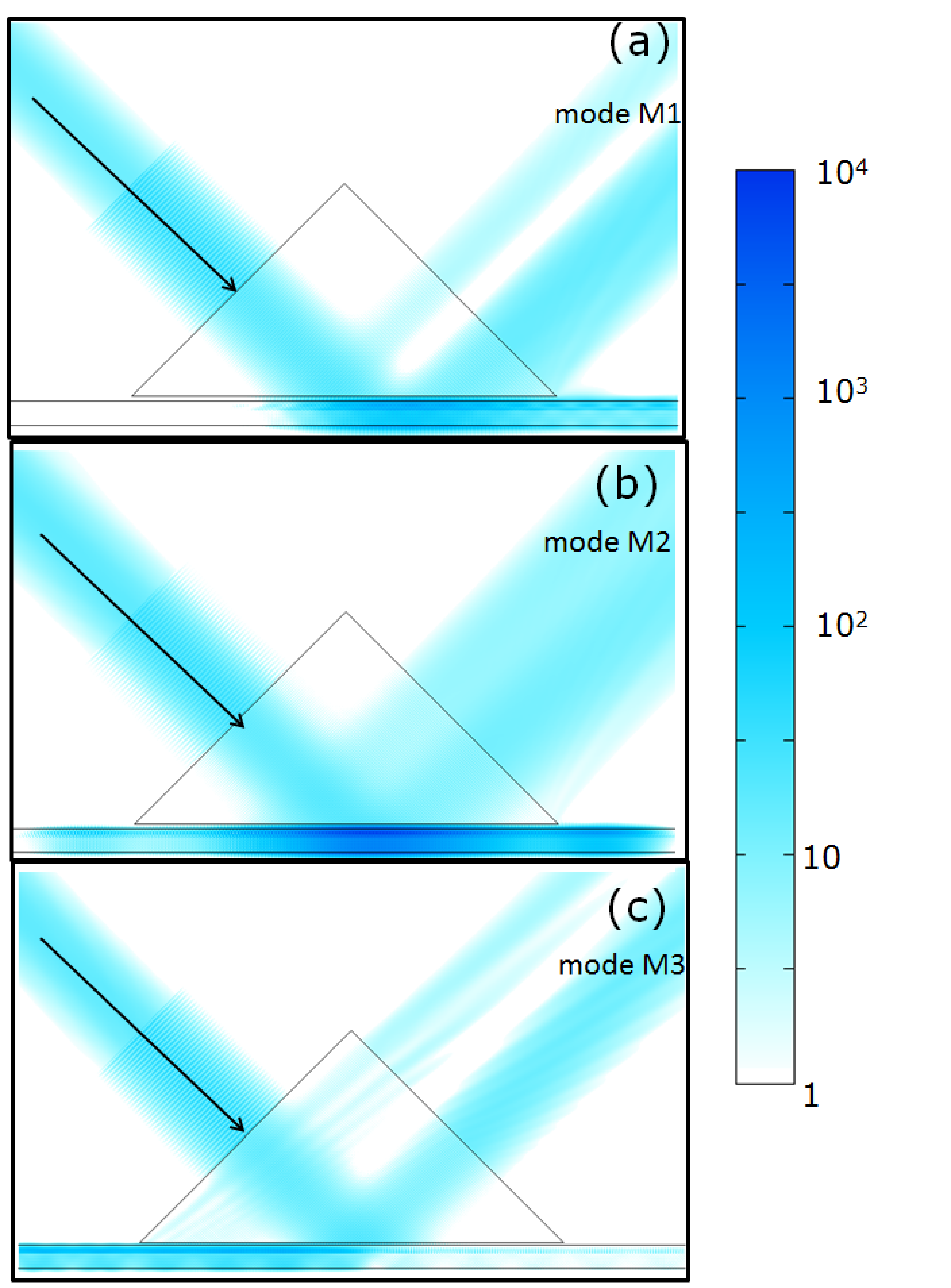}
\caption{(Color online) Same as in Fig. 2(b)-2(d) but with the ATR excitation region included. The triangle represents the ATR prism, used to excite the guided modes, while the solid lines represent the bi-waveguide limits.}
\end{center} 
\end{figure}   
\clearpage
\newpage 
\begin{figure}[!hbt]
\begin{center}   
\includegraphics[angle=0,width=8.5cm]{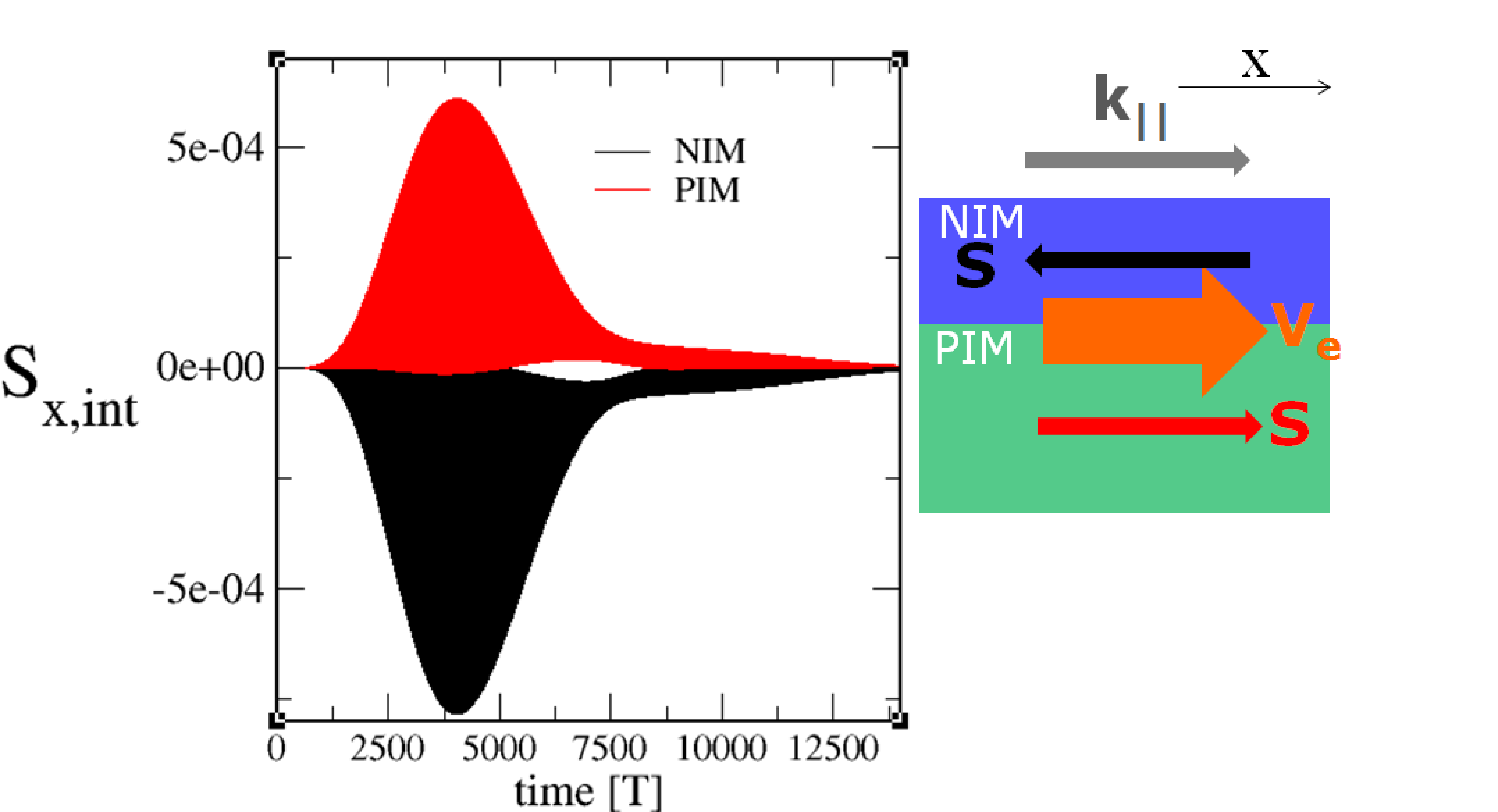}
\caption{(Color online) The  interlocking of the sub-modes within each layer of the heterostructure waveguide, observed in FDTD for mode M2 via the $x$-component of the Poynting vector monitored at the right of the ATR prism edge. The dark-colored (light-colored) line represents the result integrated over the $y$-extent of the NIM (PIM) part of the bi-waveguide. The Poynting vector is given in arbitrary units, while time is expressed in terms of the period, T of the central frequency $\omega_0$, of the pulsed signal. Clearly, the sub-modes move together towards the $+x$ direction, despite the Poynting vector showing towards the $-x$ direction in the NIM layer. [see schematics on the right of the figure indicating the wavevector ${\bf k_{||}}$ and energy velocity ${\bf v_e}$ of the composite guided mode, as well as the Poynting vector {\bf S} in each of the sub-waveguides]} 
\end{center}   
\end{figure}    
\clearpage
\newpage
\begin{figure}[!hbt]
\begin{center} 
\includegraphics[angle=0,width=8.5cm]{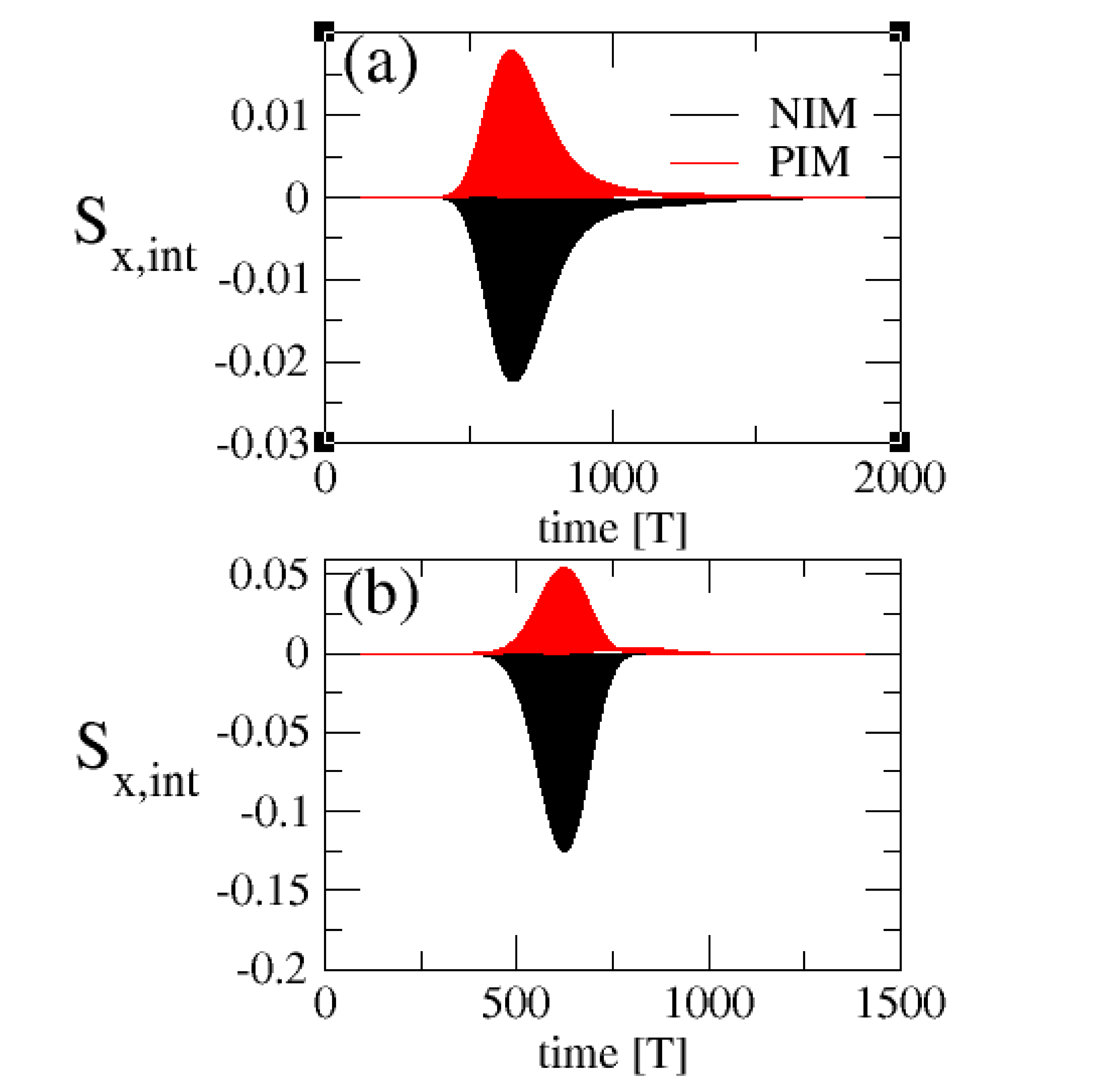}
\caption{(Color online) Same as in Fig. 4 but for the cases of mode M1 in (a), and mode M3 in (b)}
\end{center} 
\end{figure}    
\clearpage
\newpage
\begin{figure}[!hbt]
\begin{center}  
\includegraphics[angle=0,width=8.5cm]{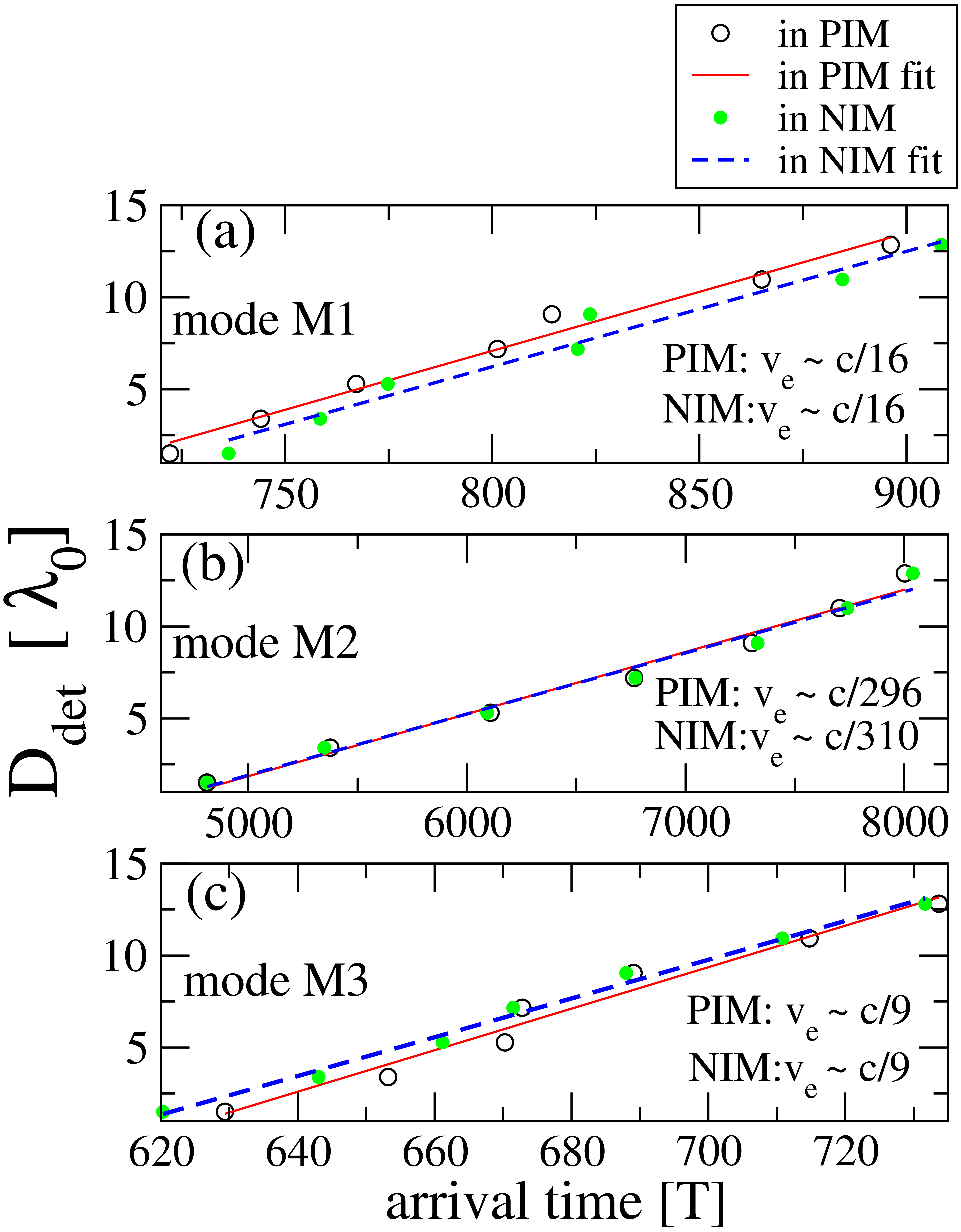}
\caption{Color online) Pulse arrival times in the NIM (PIM) part of the bi-waveguide in the different lateral detectors represented by filled (open) circles. The arrival time is given in terms of the period, while the lateral detector locations $\textrm D_{\textrm{det}}$ in terms of the wavelength that corresponds to the central frequency $\omega_0$ of the impinging pulse. The line fit in each case yields the energy velocity of propagation in each sub-waveguide.} 
\end{center} 
\end{figure}    
\clearpage
\newpage
\begin{figure}[!hbt] 
\begin{center}  
\includegraphics[angle=0,width=8.5cm]{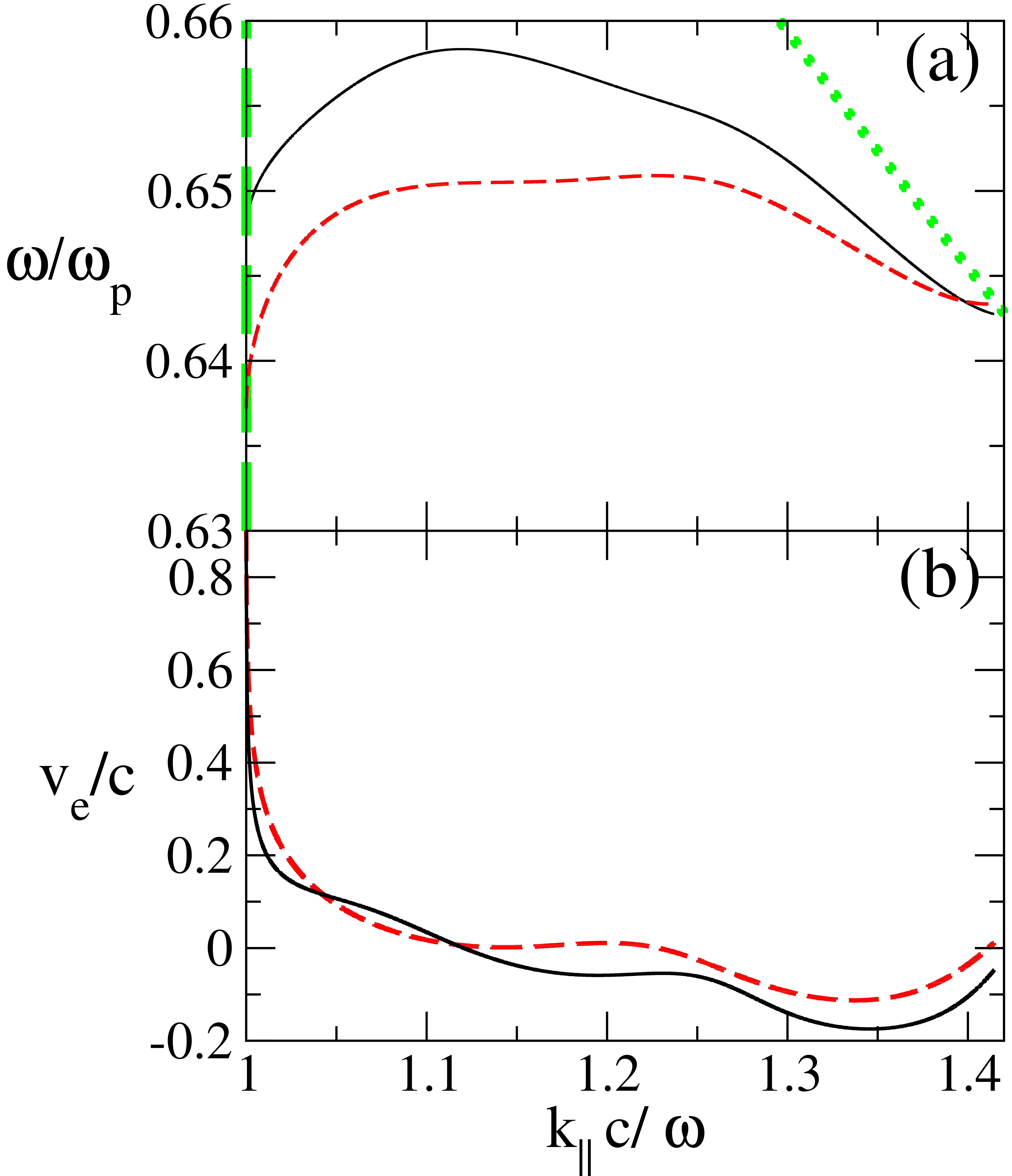}
\caption{Color online) Photonic band dispersion in (a) and energy velocity in (b) are shown with a solid line for a comparative bi-waveguide design with $\textrm d_1$=2270 nm and $\textrm d_2$=3405 nm. The green dashed and dotted lines represent the relevant lightlines as in Fig. 1. To aid the comparison, we show the respective result for the original design of Fig. 1 with dashed lines.}
\end{center} 
\end{figure}   
\clearpage
\newpage
\begin{figure}[!hbt]  
\begin{center}  
\includegraphics[angle=0,width=8.5cm]{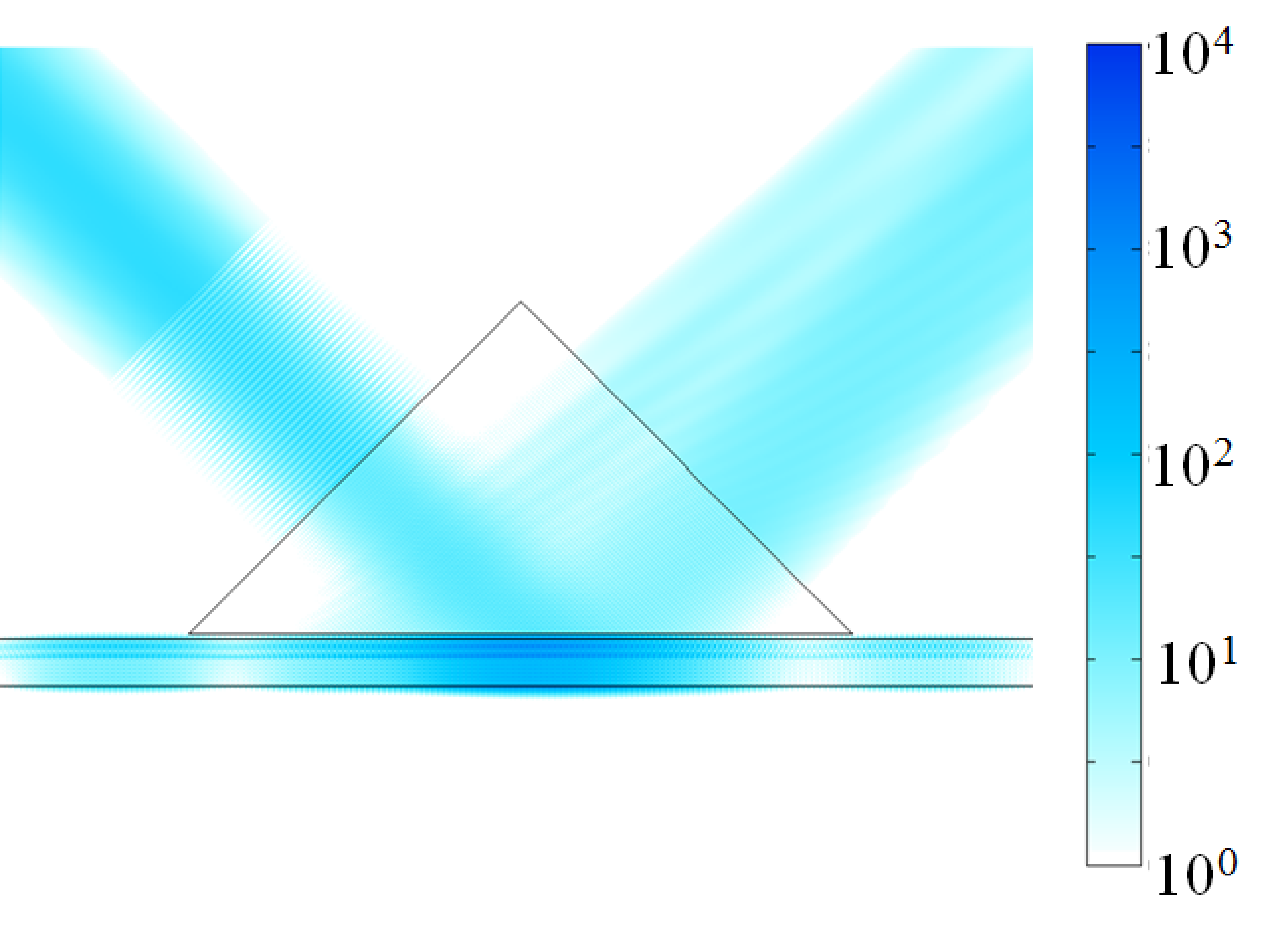}
\caption{(Color online) FDTD time-averaged intensity result, suggesting an amphoteric propagation in the comparative waveguide design of Fig. 7 for the near-frozen mode of $ck_{||}/\omega=1.12$.}
\end{center} 
\end{figure}  
\clearpage
\newpage
\begin{figure}[!hbt]
\begin{center} 
\includegraphics[angle=0,width=8.5cm]{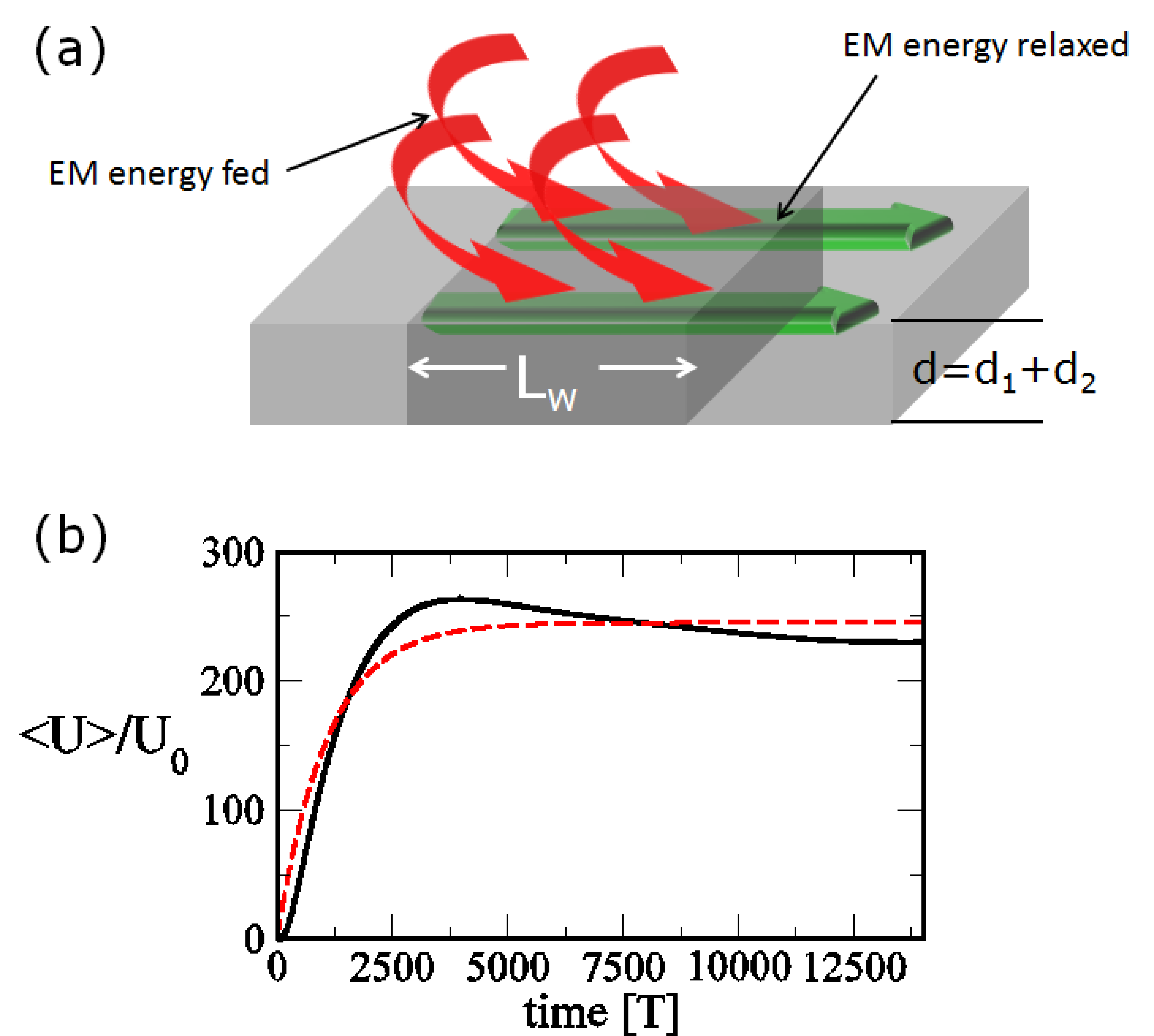}
\caption{(Color online) Energy accumulation dynamics in the bi-waveguide paradigm. (a) Feeding of EM energy from above via the ATR prism and its relaxation sideways in the bi-waveguide is depicted. (b) The solid line represents the FDTD result for the time-averaged EM energy density of the spatial average in the bi-waveguide region below the center of the ATR prism $<\textrm U>$ for mode M2, normalized with the time averaged energy density of the source $\textrm U_0$. The dotted line represents a fit from an analytical dynamical model in accordance to the picture in (a).} 
\end{center}   
\clearpage
\end{figure}
\newpage
\begin{figure}[!hbt]
\begin{center}  
\includegraphics[angle=0,width=8.5cm]{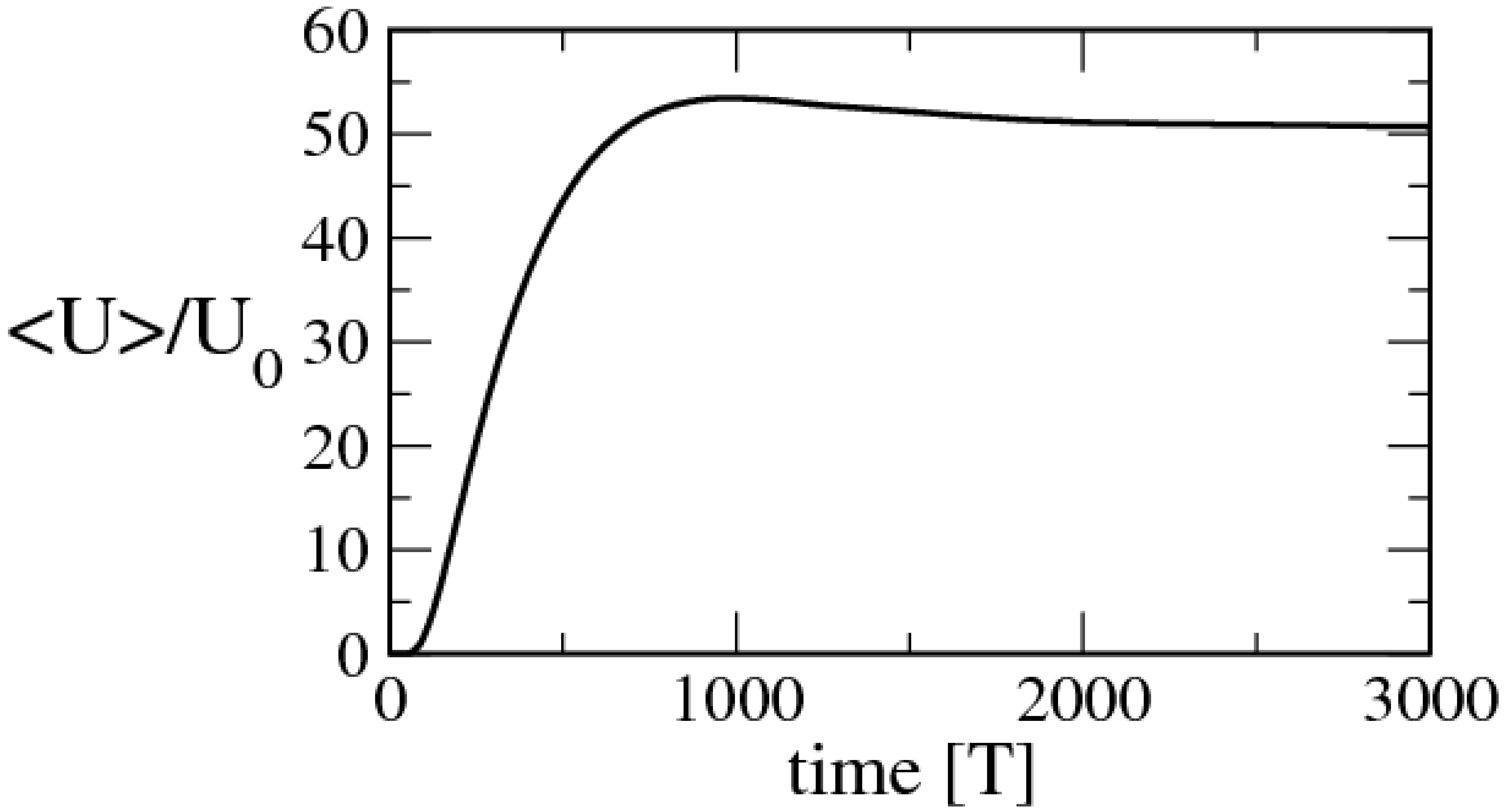}
\caption{(Color online) Same as in Fig. 9(b) but for the comparative waveguide of Fig. 7. As a result of the narrow modal bandwidth for the near-frozen mode, a smaller EM energy enhancement and quicker saturation in comparison with the design of Fig. 1 is observed.}
\end{center} 
\end{figure}    


\begin{thebibliography}{210}
\bibitem{voulaatr2} 
The distance between the prism and the waveguide $\textrm d_{\textrm{atr}}$ is adjusted so that the ATR prism is not invasive \cite{voulaback, voulaatr2}.This can be checked by confirming a small spectral offset between the response recorded by a detector above the prism, capturing the TIR beam, and the one recorded by detectors placed laterally away from the prism edge within the participating waveguides, capturing the guided mode that has coupled-in. See also Ref. \cite{voulaback}. Lateral absorbers are also placed, to reduce the effects from backreflections from the waveguide edges.
\bibitem{mono} The degree of monochromaticity of the impinging pulse,$\Delta \omega/\omega_0$, is less than $1\%$.
\bibitem{counter} We neglect here the energy relaxing in the opposite direction due to the $\Delta k_{||}$ span.
\bibitem{slowdiff} The quasi-monochromatic excitation has a sharper $\Delta \omega$ span in comparison to the pulsed signal and  thus involves a narrower part of the near-frozen light band centered around  $ck_{||}/\omega=1.183$ where theory predicts a slow-down factor of about a 100 [see Fig. 1(c)].
\end{thebibliography}
\end{document}